\documentclass[useAMS,usenatbib,usegraphicx]{mn2e}

\title[Intermittent maser flare around G353.273$+$0.641]{Intermittent maser flare around the high mass young stellar object G353.273$+$0.641 I: data \& overview}
\author[K. Motogi et al.]{K. Motogi,$^{1}$\thanks{E-mail:motogi@astro1.sci.hokudai.ac.jp}
K. Sorai,$^{1,}$$^{2}$  M. Honma$^{3,}$$^{4}$ T. Minamidani,$^{1,}$$^{2}$ T. Takekoshi,$^{1}$ 
\newauthor
K. Akiyama,$^{2,}$$^{4,}$$^{5}$ K. Tateuchi,$^{2,}$$^{6}$  K. Hosaka,$^{1}$ Y. Ohishi,$^{1}$ Y. Watanabe,$^{1,}$$^{7}$ 
\newauthor 
A. Habe,$^{1,}$$^{2}$ and H. Kobayashi$^{3,}$$^{4}$\\
$^{1}$Department of Cosmosciences, Graduate School of Science, Hokkaido University, N10 W8, Sapporo 060-0810, Japan\\
$^{2}$Department of Physics, Faculty of Science, Hokkaido University, N10 W8, Sapporo 060-0810, Japan\\
$^{3}$Department of Astronomical Science, The Graduate University for Advanced Studies, 2-21-1 Osawa, Mitaka, Tokyo 181-8588, Japan\\
$^{4}$Mizusawa VLBI Observatory, National Astronomical Observatory of Japan, 2-12 Hoshi-ga-oka, Mizusawa-ku, Oshu, Iwate 023-0861, Japan\\
$^{5}$Department of Astronomy, Graduate School of Science, The University of Tokyo, 7-3-1 Hongo, Bunkyo-ku, Tokyo 113-0033, Japan\\
$^{6}$Institute of Astronomy, University of Tokyo, Osawa 2-21-1, Mitaka, Tokyo 181-0015, Japan\\
$^{7}$Department of Physics, The University of Tokyo, 7-3-1 Hongo, Bunkyo-ku, Tokyo, 113-0033, Japan}
\begin{document}

\date{Accepted 2011 June 2. Received 2011 May 30; in original form 2011 April 29}

\pagerange{\pageref{firstpage}--\pageref{lastpage}} \pubyear{2008}

\maketitle

\label{firstpage}

\begin{abstract}
We have performed VLBI and single-dish monitoring of 22 GHz H$_{2}$O maser emission from the high mass young stellar object G353.273+0.641 
with VERA (VLBI Exploration of Radio Astrometry) and Tomakamai 11-m radio telescope. 
Two maser flares have been detected, separated almost two years. Frequent VLBI monitoring has revealed that these flare activities have been accompanied by structural change of the prominent shock front traced by H$_{2}$O maser alignments. 
We have detected only blue-shifted emissions and all maser features have been distributed within very small area of 200 $\times$ 200 au$^{2}$ in spite of wide velocity range ($>$ 100 km s$^{-1}$). 
The light curve shows notably intermittent variation and suggests that the H$_{2}$O masers in G353.273+0.641 are excited by episodic radio jet. 
The time-scale of $\sim$ 2 yr and characteristic velocity of $\sim$ 500 km s$^{-1}$ also support this interpretation. 
Two isolated velocity components of C50 (-53 $\pm$ 7 km s$^{-1}$) and C70 (-73 $\pm$ 7 km s$^{-1}$) have shown synchronised linear acceleration of the flux weighted $V_{\rmn{LSR}}$ values ($\sim$ -5 km s$^{-1}$ yr$^{-1}$) during the flare phase. 
This can be converted to the lower-limit momentum rate of 1.1 $\times$ 10$^{-3}$ M$_{\sun}$ km s$^{-1}$ yr$^{-1}$. 
Maser properties are quite similar to that of $IRAS$ 20126+4104 especially. This corroborates the previous suggestion that G353.273+0.641 is a candidate of high mass protostellar object. 
The possible pole-on geometry of disc-jet system can be suitable for direct imaging of the accretion disc in this case. 
\end{abstract}


\section{Introduction}
Astronomical masers are variable phenomenon seen in a high mass star-forming region. 
Interstellar H$_{2}$O masers ($J_{K_\rmn{a}K_\rmn{c}}$ = $6_{16}$--$5_{23}$) at 22.23508$\:$GHz, which appear at the earliest phase of formation, are highly variable in particular. 
They are often excited in strong shocks on a working surface between protostellar outflow and dense envelope (e.g., \citealt*{Furuya2000, Moscadelli2000}). 
Outflow properties such as size, velocity and morphology, 
predicted from interferometric observations of H$_{2}$O masers vary source to source 
(e.g., collimated jet \citep{Shepherd2004}, wide angle flow \citep{Motogi2008}, expanding shell \citep{Torrelles2003}, equatorial flow \citep{Trinidad2007} or combination of them \citep{Torrelles2011}). 
It is possibly related to a number of factors such as stellar mass, evolutional stage, geometry of surrounding envelope and driving mechanism of host outflow. 
Several single-dish based studies have reported sub-yr scale maser flares in some cases (e.g., \citealt{Felli2007}). 
This time-sale is comparable with or even shorter than that of episodic radio jet, which traces a root of protostellar outflow (e.g., \citealt{Anglada1996} and reference therein), 
and indicates that some portion of H$_{2}$O masers contains valuable information about an innermost outflow. 
Exact relation between an outflow activity and maser variability is still unclear, 
but once it is set up, H$_{2}$O maser will be an excellent tool to survey dynamic outflow activities in small scale, 
since its bright emission allows us easy and frequent {\bf monitoring even with} a small size radio telescope.

In this paper, we report on the new VLBI and single-dish study for 
intermittent flare activities of H$_{2}$O masers in G353.273+0.641 (hereafter G353). 
G353 is a strong 22 GHz H$_{2}$O maser site in the southern high mass star-forming region NGC6357 \citep{Sakellis1984}. 
The source distance is 1.7 kpc from the sun \citep{Neckel1978}. 
Multi-epoch ATCA observation has been reported in \citet{Caswell2008} (hereafter CP08). 
Class II CH$_{3}$OH maser emission ($J_k$ = 5$_{1}$--6$_{0}$ A$^{+}$) at 6.668519 GHz is also associated with this source (CP08), 
and hence, the host young stellar object (YSO) is identified as a high mass YSO (e.g., \citealt{Minier2003}). 
CP08 has also suggested that G353 is still in the pre-ultra compact (UC) H\,{\small\bf II} region phase, i.e., high mass protostellar phase, 
based on the absence of any detectable OH masers (e.g., \citealt{Caswell1997,Breen2010}). 

G353 has been classified as dominant blue-shifted H$_{2}$O maser in CP08. That is, almost all flux is concentrated on blue-shifted emission, 
despite of very broad velocity range of $\pm$ 100 km s$^{-1}$ with respect to the systemic velocity of -5 km s$^{-1}$. 
They have argued that this type of masers can be caused by well-collimated jet aligned close to the line of sight. 
There are a few maser sources which show similar blue-shift dominance (e.g., \citealt{Caswell2004}; CP08; \citealt{Caswell2010}). 
Some of them suggest acceleration of outflowing materials in the jet. 
The statistical analysis in \citet{Caswell2010} indicates that such a blue-shift dominance 
is a characteristic of H$_{2}$O masers at the earliest evolutionary stage of star-formation. 

Although the signature of yr-scale variability of the H$_{2}$O maser has been pointed out by CP08, 
their small dataset is still insufficient for detailed discussion. 
Our VLBI and single-dish monitoring with VERA (VLBI Exploration of Radio Astrometry) and the Hokkaido University Tomakomai 11-m radio telescope in Japan, 
which has been started from November 2008, has provided more tight-interval dataset and detected intermittent flare activities accompanied by spatial changes of the shock front traced by maser distribution. 

\begin{table*}
  \caption{Summary of VERA Obsevations}\label{tb:obs}
   \begin{tabular}{cccccccc}\hline
   Epoch&Date& Synthesised Beam & PA & 1-$\sigma$ Noise$^{a}$ & DN$_{p}$$^{b}$ / DN$_{s}$ &Positional Error & Comments \rule[0mm]{0mm}{4mm}\\
       && (mas $\times$ mas) & ($^{\circ}$) & (Jy beam$^{-1}$) &($\sigma$) & (mas $\times$ mas) &for $T_{\mathrm{sys}}$\\ \hline
1&2008/7/1 & 3.16 $\times$ 0.74 & -17.1 & 0.48  & 13 / 117& 0.15 $\times$ 0.49 &-- \\
2&2008/11/18 & 3.38 $\times$ 0.78 & -21.7 &0.30  & 6 /49& 0.92 $\times$ 2.15 &$\sim$ 1000 K at OG\\
3&2009/2/6 & 2.63 $\times$ 0.91 & -14.1 &0.37  & 14 / 351& 0.06 $\times$ 0.24 & --\\
4&2009/5/18 & 2.57 $\times$ 0.88 & -13.2 & 0.30  & 16 / 1122& 0.04 $\times$ 0.16 & \\
5&2009/9/16 & 2.86 $\times$ 0.82 & -12.3 & 0.37 & 10 / 19& 0.11 $\times$ 0.48 & --\\
6&2009/11/5 & 2.79 $\times$ 0.79 & -13.3 & 0.16  & 15 / 338& 0.23 $\times$ 0.98 & -- \\
7&2010/1/21 & 3.15 $\times$ 0.85 & -19.5 & 0.21  & 15 / 72& 0.09 $\times$ 0.25 & $\sim$ 2000 K at OG\\
8&2010/4/21 & 2.59 $\times$ 0.75 & -14.1 & 0.40  & 12 / 184& 0.10 $\times$ 0.41 &$\sim$ 1500 K at IS\\ \hline
\multicolumn {8} {l} {$^a$ Typical value in self-calibrated images.}\\
\multicolumn {8} {l} {$^b$ The dynamic ranges of the brightest spot in phase-referenced (DN$_{p}$) and self-calibrated image (DN$_{s}$).}\\
\multicolumn {8} {l} {$^c$ IR, IS, OG:Iriki, Ishigaki, Ogasawara station, respectively.} \\
     \end{tabular}
\end{table*}

\section{Observations and data reduction}
\subsection{VLBI observations}
VLBI monitoring of H$_{2}$O masers with VERA \citep{Kobayashi2008} has been performed eight epochs from November 2008 to April 2010 and is still on-going. 
Each observation was made in VERA's dual-beam mode in which targeted maser source and phase calibrator were observed simultaneously (\citealt{Kawaguchi2000, Honma2003}). 

Left-handed circular polarised signals were quantised at 2-bit sampling and filtered with the VERA digital filter unit \citep{Iguchi2005}, 
after that, data were recorded onto magnetic tapes at a data rate of 1024 Mbps. 
The data correlation was performed with the Mitaka FX correlator \citep{Shibata1998}. 
Total bandwidths assigned for calibrators and the maser are 240 MHz (16 IFs) and 8 MHz (1 IF), respectively. 
Whole velocity coverage of 108 km s$^{-1}$ were divided into 512 spectral channels yielding final spectral resolution of 0.21 km s$^{-1}$. 

Special calibration methods for VERA's dual-beam instrument have already been described in \citet{Motogi2011}. 
We used J1717-3342 \citep{Petrov2006} as a phase calibrator. 
This source is separated from G353 by 1\degr.83 at a position angle of -108\degr$\:$ east of north. 
J1717-3342 has been enough bright ($\sim$ 0.7 Jy) and shown no significant structure during two years. 
We also observed NRAO530 (=J1733-1304: \citealt{Ma1998}) as a delay and bandpass calibrator every 120 minutes. 
We have not observed any flux calibrator and antenna gains have been corrected 
based on measured system noise temperatures ($T_{\rmn{sys}}$) of each station. 
Typical $T_{\rmn{sys}}$ values were $\sim$ 350 K for Mizusawa and Iriki stations and $\sim$ 500 K for Ogasawara and Ishigaki stations in the case of normal weather condition. 
Total on-source time is about 2.5 hour in the first five epochs and 5 hour in the last three epochs. 
We note that the observations were time-shared with another maser -- calibrator pair, G5.89-0.39 -- J1755-2232, in the first five epochs \citep{Motogi2011}. 
This causes relatively short on-source time in the former five epochs, but total hour angle range (i.e., achieved $UV$ -- coverage) is almost same as the latter three epochs. 

Data reduction was carried out using NRAO Astronomical Imaging Processing System (AIPS) package. 
We adopted standard reduction procedure for VERA data (See \citealt{Motogi2011} in details), where 
we have first determined absolute coordinates of several maser spots in 2-beam phase referencing analysis, 
and then, performed self-calibration using the brightest maser spot in each epoch. 
We finally searched all maser spots within 1\arcsec.5$\times$1\arcsec.5 area with 7-$\sigma$ detection limit. 
Here, the term ’maser spot’ indicates a maser component seen in a single velocity channel, and hereafter, 
the term ’maser feature’ means physical gas clump which is consist of several maser spots detected in successive velocity channels and closely located each other. 

Table 1 summarised all the VERA observations. It contains observing dates, synthesised beam sizes in milli-arcsecond (mas), 
beam position angles (PA) east of north, typical noise levels and dynamic ranges of synthesised images for the brightest maser spot, estimated errors of absolute positions (see section 3.1) and brief comments, 
if any, about $T_{\rmn{sys}}$ value.  

\subsection{Single-dish observations}
Single-dish monitoring has been done using the Hokkaido University Tomakomai 11-m radio telescope \citep{Sorai2008}.  
Observing interval is about 1 -- 3 times per month except for the receiver maintenance session during July to October 2009. 
The beam width (FWHM) and aperture efficiency is 4\arcmin.2 and 49 \% at 22 GHz, respectively. 
These values have been slightly improved compared with that mentioned in \citet{Sorai2008} because of the upgrade in optics. 
The pointing accuracy is better than 30\arcsec$\:$ in each of azimuth and elevation direction. 
Observations were performed in the position switching method. 
We received single left circular polarisation with 16-MHz bandwidth (216 km s$^{-1}$ in velocity) and it was divided into 2048 spectral channels. 
An intensity scale was calibrated by a chopper wheel method every 10 minutes. 

Observed data were reduced with the NEWSTAR software package developed in Nobeyama Radio Observatory. 
Initial 2048 spectral channels were binned up every two channels, in order to obtain the same spectral resolution as VERA data. 
Typical rms noise level is $\sim$ 3 Jy in winter and increases up to 6 -- 25 Jy in the summer season, 

\section{Results}
\subsection{Spatial and velocity distribution}

\begin{figure*}
\includegraphics*[trim=0 25 0 20,scale=0.75]{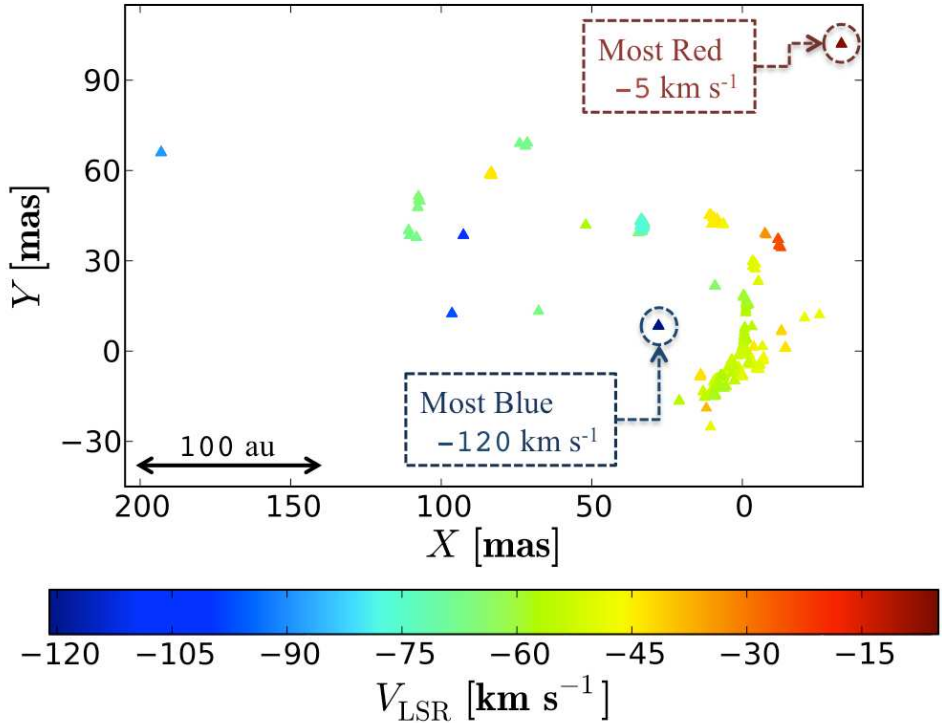}\\
\caption{Spatial and velocity distribution of all maser spots in G353. 
Colour triangles indicate detected maser spots and their $V_{\rmn{LSR}}$ values. 
Coordinate origin is ($\alpha$,$\delta$)$_{\rmn{J2000.0}}$ = ($17^{\rmn{h}}26^{\rmn{m}}01^{\rmn{s}}.5883$, -34\degr 15\arcmin 14\arcsec.905) 
which is the position of the brightest maser spot in second VLBI epoch.}
\end{figure*}

Figure 1 shows the spatial and velocity distribution of H$_{2}$O maser spots in G353. 
All detected maser spots in eight epochs are exhibited in this figure, since several maser features have significant velocity width and spatial elongation. 
Detailed parameters of identified maser features are listed in table 2. 
The $X$, $Y$ axes and colour show the right ascension offset ($\Delta\alpha$$\times$cos$\delta$), declination offset ($\Delta\delta$) and $V_{\rmn{LSR}}$, respectively.  
The coordinate origin is ($\alpha$,$\delta$)$_{\rmn{J2000.0}}$ = ($17^{\rmn{h}}26^{\rmn{m}}01^{\rmn{s}}.5883$, -34\degr 15\arcmin 14\arcsec.905), 
which is the absolute position of the brightest maser spot in the second {\bf VLBI} epoch. 
This position is consistent with CP08 position within the range of errors. 

Estimated errors of our position measurements are listed in table 1. 
We have evaluated three error sources, i.e., thermal noise in the phase-referenced images, 
errors of baseline vectors and atmospheric phase fluctuations in the same way as \citet{Nakagawa2008}.  
The atmospheric zenith delay residuals ($\sim$ 0.7 cm on average) are determined by the image optimizing method (\citealt{Honma2007}, 2008). 
The largest atmospheric residuals ($\sim$ 3 cm) was found in the second VLBI epoch, where estimated error reached 1 mas in $X$ and 2 mas in $Y$ directions. 
The relative position uncertainty between each maser spot is, on the other hand, typically 30 $\mu$as in $X$ and 100 $\mu$as in $Y$ directions, 
based on thermal noise in the self-calibrated images. 

All maser spots have been found within 200 mas $\times$ 100 mas area. 
This corresponds to 350 au $\times$ 200 au at the source distance of 1.7 kpc. 
We have detected successive maser alignment, which looks like a filament, in the central 50 mas $\times$ 50 mas region. 
This filamentary structure is pronounced only in the flare-up phase tracing shock front (see next subsection). 

Total $V_{\rmn{LSR}}$ coverage of maser emissions in VERA observations is -121 to -5 km s$^{-1}$, and almost same as the single-dish data. 
This is well consistent with the blue-shifted part of CP08 spectrum. 
The red end of -5 km s$^{-1}$ is same as the systemic velocity. 
Extremely red (+87 km s$^{-1}$) component reported in CP08 is out of our observing band. 
Large portion of flux is concentrated in the range of -80 to -40 km s$^{-1}$. 

Spatially, there is no clear gradient of $V_{\rmn{LSR}}$ for features in this dominant velocity range. 
The features out of this range roughly split into east (blue) and west (red) part of G353. 
The most blue and red components are separated only 100 mas (200 au) along SE -- NW direction.  
The wide range of velocity within a very compact region readily dismisses the possibility of disc maser which has been proposed in a few cases (e.g., \citealt{Imai2006, Trinidad2009}). 
It also indicates relatively pole-on geometry of the disc-jet system as suggested in CP08. 
Expected young age of the driving source may also be related to such a concentrated distribution, 
however, observed proper motions (see next section) comparable to the $V_{\rmn{LSR}}$ values still support moderately pole-on geometry even in this case.  

\subsection{Flare activity}
We have detected strong maser flare in the second VLBI epoch. 
Time variation of single-dish spectrum after the flare is displayed in figure 2. 
There are two velocity components which show significant brightening. 
One is the component of -53 $\pm$ 7 km s$^{-1}$ (hereafter C50) and another is the component of -73 $\pm$ 7 km s$^{-1}$ (hereafter C70). 
Almost all emissions disappeared in the quiet phase. 
Figure 3 presents the light curves for both flared components during 950 days from Jan 1st, 2008. 
The vertical axis indicates peak flux density for single-dish data, and sum of the flux densities of maser spots detected in the peak velocity channel for VLBI data. 
Downward arrows show 3-$\sigma$ upper limit for C70. 
The plotted data correspond to column 5 of table 3. 

Only C50 had been flared at first and C70 started to increase its flux about 100 days after observed C50 peak. 
Sign of second flare is detected only for C50 after the 8th VLBI epoch (day 841). 
The systematic increase of the 3-$\sigma$ upper limits in the latter part of C70 is not a signature of second flare, but because of bad weather condition in summer season. 
A variation pattern of H$_{2}$O maser is, in general, combination of long-term, global pattern and short-term fluctuation (e.g., \citealt{Felli2007}). 
Smoothly varying light curve indicates that global variation pattern is dominant in C50 rather than random fluctuation during our monitoring. 
Although large dip was caught in day 473 for C70, same global pattern was identifiable, i.e., the flare-up continued for $\sim$ 150 days and rapid decaying. 
The dip in C70 is roughly factor of three weaker than the peak flux density and explainable by short-term fluctuation. 

Figure 4 presents time variation of the prominent part of the maser alignment that contains most flux of C50. 
This successive alignment had been appeared after the first flare. 
The maximum spatial scale of the alignment is roughly 50 mas ($\sim$ 90 au) that is almost typical value for an individual maser filament in an outflow shock (e.g., \citealt*{Elitzur1989}; \citealt{Motogi2008, Torrelles2011}). 
A number written in upper right part of each panel shows relative day from the 1st VLBI epoch (day 183). 
The alignment was initially vertical and dramatically changed to be horizontal during the flare phase (2nd -- 4th epoch). 
All maser features in this region completely disappeared in the 7th VLBI epoch. 
Another maser alignment appeared in the final VLBI epoch, tracing the second flare. 
 
Because of non-negligible accelerations of $V_{\rmn{LSR}}$, 
the identification of individual maser feature in these alignments is difficult (see section 3.3). 
We, thus, estimated averaged proper motions of these alignment, where 
we averaged the positions of all maser spots in each alignment and calculated $\mu_{X}$ and $\mu_{Y}$ for vertical and horizontal alignments, respectively. 
Here, $\mu_{X}$ and $\mu_{Y}$ show the linear proper motion in each directions. 
Such a motion perpendicular to a maser alignment should trace bulk motion of shock front, if we 
consider the standard model that H$_{2}$O masers are excited in a thin sheet of dense gas compressed by outflow shock \citep{Elitzur1989}. 
Estimated $\mu_{X}$ and $\mu_{Y}$ are 7.8 mas yr$^{-1}$ from east to west and 7.6 mas yr$^{-1}$ from north to south, respectively. 
These averaged proper motions correspond to $\sim$ 60 km s$^{-1}$ at 1.7 kpc and comparable with their $V_{\rmn{LSR}}$ values of -50 -- -70 km s$^{-1}$. 

\setcounter{table}{1}
\begin{table*}
\centering 
\begin{minipage}{120mm}
\caption{Detected Maser Features}
\begin{tabular}{cccccc}\hline
Epoch & $V_{\rmn{LSR}}$ & \multicolumn{2}{c}{Offset$^a$ (mas)} & Total Width$^{b}$ & Integrated Flux  \\
& (km s$^{-1}$)&  $X$ & $Y$ & (km s$^{-1}$) & (Jy km s$^{-1}$) \\ \hline
1 & -43.57 & 9.65 (0.01) & 42.34 (0.02) & 1.69 & 13.01 (0.44) \\ 
 & -44.21 & 10.14 (0.02) & 42.25 (0.04) & 0.42 & 0.61 (0.05) \\ 
 & -46.52 & -0.31 (0.03) & -0.90 (0.07) & 1.05 & 0.89 (0.09) \\ 
 & -57.9 & -1.65 (0.01) & 15.20 (0.01) & 1.48 & 6.19 (0.14) \\ 
 & -63.38 & 9.05 (0.00) & 21.69 (0.01) & 1.90 & 7.25 (0.13) \\ 
 & -66.75 & 34.43 (0.01) & 39.46 (0.02) & 1.48 & 3.85 (0.13) \\ 
 & -88.03 & 193.03 (0.01) & 66.06 (0.01) & 1.26 & 2.70 (0.05) \\ \hline
2 & -42.69 & 13.93 (0.02) & -8.08 (0.04) & 1.05 & 3.44 (0.16) \\ 
 & -43.95 & 10.79 (0.01) & 45.17 (0.02) & 1.48 & 5.08 (0.12) \\ 
 & -48.8 & 0.47 (0.01) & -0.95 (0.02) & 3.37 & 134.99 (1.89) \\ 
 & -51.96 & 4.64 (0.00) & -5.51 (0.01) & 5.48 & 445.00 (4.09) \\ 
 & -49.85 & -0.28 (0.02) & 1.14 (0.03) & 0.63 & 3.78 (0.12) \\ 
 & -51.54 & 0.16 (0.01) & 0.38 (0.02) & 2.53 & 87.12 (1.28) \\ 
 & -52.38 & 3.66 (0.01) & -3.14 (0.01) & 0.63 & 0.92 (0.05) \\ 
 & -51.96 & 5.88 (0.00) & -7.99 (0.01) & 0.63 & 0.99 (0.06) \\ 
 & -51.96 & -1.10 (0.03) & 3.60 (0.03) & 0.21 & 0.10 (0.02) \\ 
 & -52.17 & -0.07 (0.02) & 1.32 (0.04) & 0.21 & 0.61 (0.03) \\ 
 & -52.38 & -1.12 (0.04) & 4.38 (0.05) & 0.21 & 0.07 (0.01) \\ 
 & -52.59 & -0.22 (0.02) & 2.10 (0.05) & 0.63 & 1.41 (0.06) \\ 
 & -53.43 & -0.55 (0.03) & 3.95 (0.06) & 2.32 & 6.29 (0.26) \\ 
 & -56.17 & -0.79 (0.05) & 7.79 (0.12) & 0.21 & 0.01 (0.00) \\ 
 & -56.38 & -1.19 (0.02) & 14.43 (0.08) & 1.05 & 0.29 (0.01) \\ 
 & -58.7 & -0.39 (0.01) & 18.34 (0.02) & 1.9 & 2.35 (0.04) \\ 
 & -63.76 & 107.57 (0.06) & 51.48 (0.11) & 0.42 & 0.03 (0.00) \\ 
 & -66.29 & 110.83 (0.04) & 40.11 (0.08) & 0.63 & 0.04 (0.00) \\ 
 & -67.97 & 32.99 (0.01) & 43.13 (0.02) & 1.26 & 0.63 (0.01) \\ 
 & -69.87 & 33.76 (0.01) & 43.07 (0.02) & 1.48 & 0.81 (0.02) \\ 
 & -72.18 & 33.52 (0.02) & 43.85 (0.04) & 0.84 & 0.11 (0.01) \\ \hline
3 & -5.72 & -32.93 (0.02) & 102.08 (0.04) & 0.84 & 0.94 (0.07) \\
 & -23.63 & -11.92 (0.01) & 37.09 (0.02) & 1.05 & 1.20 (0.05) \\ 
 & -32.68 & -7.70 (0.05) & 38.91 (0.06) & 0.63 & 0.15 (0.02) \\ 
 & -41.53 & -12.94 (0.05) & 6.67 (0.13) & 0.21 & 0.02 (0.00) \\ 
 & -43.43 & -3.80 (0.04) & 1.39 (0.09) & 1.05 & 0.29 (0.03) \\ 
 & -44.06 & 9.46 (0.01) & 44.27 (0.01) & 1.48 & 2.04 (0.04) \\ 
 & -45.32 & 6.37 (0.01) & 42.16 (0.03) & 1.69 & 0.39 (0.02) \\ 
 & -48.06 & -3.76 (0.01) & 29.58 (0.01) & 2.32 & 2.03 (0.04) \\ 
 & -48.9 & -2.00 (0.03) & -1.15 (0.06) & 1.26 & 0.38 (0.02) \\ 
 & -51.85 & -2.03 (0.01) & -0.87 (0.03) & 2.95 & 6.37 (0.26) \\ 
 & -50.38 & 1.29 (0.01) & -5.55 (0.02) & 1.26 & 1.27 (0.03) \\ 
 & -53.12 & 3.39 (0.00) & -7.28 (0.00) & 3.58 & 137.82 (0.40) \\ 
 & -51.64 & 8.74 (0.02) & -9.32 (0.02) & 2.53 & 3.79 (0.16) \\ 
 & -53.33 & 2.71 (0.01) & -4.75 (0.00) & 0.63 & 0.05 (0.01) \\ 
 & -54.59 & 6.44 (0.01) & -8.69 (0.02) & 1.9 & 1.82 (0.05) \\ 
 & -56.7 & 7.06 (0.02) & -8.31 (0.02) & 1.47 & 0.28 (0.01) \\ 
 & -55.22 & -1.41 (0.03) & 4.03 (0.05) & 0.84 & 0.08 (0.01) \\ 
 & -55.43 & -3.14 (0.04) & 8.13 (0.10) & 0.21 & 0.01 (0.00) \\ 
 & -58.38 & -1.87 (0.01) & 15.95 (0.02) & 1.48 & 0.17 (0.01) \\ 
 & -64.49 & 107.28 (0.01) & 50.10 (0.01) & 2.11 & 0.39 (0.01) \\ 
 & -66.81 & 108.18 (0.02) & 37.78 (0.04) & 0.84 & 0.03 (0.00) \\ 
 & -66.81 & 67.69 (0.04) & 13.31 (0.09) & 0.42 & 0.00 (0.00) \\ 
 & -67.44 & 110.58 (0.03) & 38.34 (0.06) & 0.42 & 0.00 (0.00) \\ 
 & -70.39 & 32.54 (0.00) & 42.4 (0.00) & 2.74 & 1.68 (0.01) \\ 
 & -72.71 & 32.79 (0.00) & 42.81 (0.00) & 2.11 & 1.40 (0.01) \\ 
 & -73.76 & 33.00 (0.00) & 42.92 (0.01) & 1.90 & 0.58 (0.01) \\ \hline
 4 & -23.91 & -12.31 (0.01) & 34.98 (0.02) & 1.47 & 6.83 (0.18) \\ 
 & -43.93 & 83.35 (0.03) & 59.54 (0.06) & 0.21 & 0.09 (0.01) \\ 
 & -44.35 & 9.10 (0.06) & 42.72 (0.11) & 0.21 & 0.03 (0.01) \\ 
 & -44.35 & -0.04 (0.01) & -8.35 (0.02) & 0.84 & 1.58 (0.04) \\ \hline
 \end{tabular}
\end{minipage}
\end{table*}
\begin{table*}
\centering 
\begin{minipage}{120mm}
\contcaption{}
\begin{tabular}{cccccc}\hline
Epoch & $V_{\rmn{LSR}}$ & \multicolumn{2}{c}{Offset$^a$ (mas)} & Total Width$^{b}$ & Integrated Flux  \\
& (km s$^{-1}$)&  $X$ & $Y$ & (km s$^{-1}$) & (Jy km s$^{-1}$) \\ \hline
 & -44.56 & 0.10 (0.01) & -8.42 (0.02) & 0.63 & 0.73 (0.02) \\ 
 & -48.56 & -2.83 (0.01) & -3.57 (0.03) & 1.48 & 2.76 (0.08) \\ 
 & -48.77 & -3.86 (0.02) & 27.88 (0.04) & 1.47 & 0.81 (0.05) \\ 
 & -49.61 & -0.01 (0.03) & -8.01 (0.05) & 0.84 & 0.77 (0.03) \\ 
 & -52.77 & 10.05 (0.01) & -12.93 (0.02) & 1.90 & 2.33 (0.07) \\ 
 & -51.72 & 13.01 (0.02) & -13.39 (0.05) & 1.05 & 0.34 (0.02) \\ 
 & -53.83 & 2.62 (0.01) & -9.81 (0.03) & 1.69 & 2.06 (0.08) \\ 
 & -53.2 & 9.93 (0.01) & -12.67 (0.01) & 1.05 & 1.36 (0.03) \\ 
 & -55.09 & 5.47 (0.01) & -11.64 (0.02) & 1.05 & 1.32 (0.04) \\ 
 & -54.88 & 7.36 (0.03) & -12.12 (0.03) & 1.26 & 0.94 (0.06) \\ 
 & -56.14 & 7.10 (0.02) & -11.70 (0.02) & 1.05 & 0.35 (0.02) \\ 
 & -56.14 & 8.04 (0.03) & -11.61 (0.03) & 1.69 & 0.46 (0.03) \\ 
 & -65.62 & 107.80 (0.02) & 47.85 (0.05) & 0.63 & 0.05 (0.00) \\ 
 & -69.84 & 33.25 (0.02) & 40.05 (0.05) & 0.63 & 0.05 (0.00) \\ 
 & -71.94 & 32.86 (0.00) & 41.11 (0.01) & 2.95 & 3.10 (0.04) \\ 
 & -74.47 & 33.31 (0.00) & 41.60 (0.00) & 2.74 & 14.07 (0.03) \\ 
 & -95.96 & 96.44 (0.01) & 12.48 (0.03) & 0.84 & 0.08 (0.00) \\ 
 & -120.18 & 27.84 (0.01) & 8.38 (0.02) & 1.69 & 0.39 (0.01) \\ \hline
5 & -43.74 & 83.18 (0.03) & 58.33 (0.08) & 0.42 & 1.01 (0.13) \\ 
 & -49.43 & -4.83 (0.01) & -5.81 (0.02) & 2.11 & 11.75 (0.34) \\ 
 & -54.7 & 12.57 (0.02) & -15.34 (0.06) & 0.42 & 0.24 (0.03) \\ 
 & -56.6 & 9.12 (0.03) & -14.84 (0.05) & 1.05 & 0.77 (0.07) \\ 
 & -75.14 & 33.00 (0.01) & 41.03 (0.02) & 1.69 & 4.67 (0.16) \\ \hline
 6 & -24.13 & -12.84 (0.04) & 34.28 (0.14) & 0.21 & 0.25 (0.05) \\ 
 & -42.88 & -14.26 (0.01) & 1.07 (0.06) & 0.84 & 1.36 (0.09) \\ 
 & -43.72 & 83.85 (0.01) & 58.7 (0.03) & 0.63 & 0.41 (0.03) \\ 
 & -45.41 & -5.21 (0.04) & -5.68 (0.08) & 0.21 & 0.04 (0.00) \\ 
 & -47.94 & 10.58 (0.06) & -25.2 (0.09) & 0.21 & 0.05 (0.01) \\ 
 & -49.41 & -4.78 (0.00) & -5.68 (0.01) & 2.53 & 7.10 (0.09) \\ 
 & -48.78 & -6.67 (0.01) & 1.54 (0.04) & 0.42 & 0.08 (0.01) \\ 
 & -57.42 & 9.65 (0.01) & -14.76 (0.02) & 1.26 & 0.54 (0.02) \\ 
 & -67.32 & 72.03 (0.01) & 68.13 (0.02) & 0.84 & 0.23 (0.01) \\ 
 & -75.96 & 33.86 (0.00) & 41.62 (0.00) & 1.90 & 4.07 (0.04) \\ 
 & -107.35 & 92.64 (0.01) & 38.53 (0.02) & 1.05 & 0.39 (0.01) \\ \hline
7 & -50.31 & -5.23 (0.05) & 23.23 (0.11) & 0.63 & 0.94 (0.15) \\ 
 & -56.63 & 52.02 (0.03) & 41.85 (0.06) & 0.63 & 0.74 (0.07) \\ 
 & -67.16 & 71.42 (0.01) & 69.19 (0.02) & 1.05 & 3.96 (0.09) \\ 
 & -77.27 & 33.30 (0.03) & 42.95 (0.07) & 0.63 & 0.22 (0.02) \\ \hline
8 & -38.38 & 11.95 (0.02) & -18.89 (0.06) & 0.21 & 0.96 (0.09) \\ 
 & -47.22 & -6.98 (0.01) & -2.75 (0.03) & 1.26 & 6.14 (0.26) \\ 
 & -47.43 & -25.62 (0.02) & 12.09 (0.13) & 0.21 & 0.14 (0.03) \\ 
 & -47.43 & -20.70 (0.03) & 11.09 (0.13) & 0.21 & 0.11 (0.03) \\ 
 & -48.49 & -5.75 (0.00) & -4.57 (0.01) & 2.11 & 24.19 (0.30) \\ 
 & -51.65 & -2.10 (0.01) & -3.84 (0.03) & 1.26 & 1.15 (0.05) \\ 
 & -55.86 & 20.96 (0.04) & -16.42 (0.08) & 0.42 & 0.21 (0.02) \\ 
 & -58.39 & 5.95 (0.03) & -12.31 (0.10) & 0.21 & 0.03 (0.00) \\ 
 & -67.02 & 73.96 (0.02) & 69.02 (0.06) & 0.42 & 0.08 (0.01) \\ \hline
 \multicolumn{6}{l}{$^a$ Relative offsets from ($\alpha$,$\delta$)$_{\rmn{J2000.0}}$ = ($17^{\rmn{h}}26^{\rmn{m}}01^{\rmn{s}}.5883$, -34\degr 15\arcmin 14\arcsec.905).}\\
 \multicolumn{6}{l}{$^b$ Full width at zero intensity (FWZI) for each maser feature.}\\
 \multicolumn{6}{l}{All parenthetic values indicate associated errors.}\\
\end{tabular}
\end{minipage}
\end{table*}%

We note that these averaged motions are absolute motions respect to the referenced calibrator, and hence, 
include both of the solar motion relative to the LSR and the systemic motion of G353. 
The solar motion has already been subtracted from averaged motions above. 
We used the solar motion estimated from Hipparcos data (0.3  mas yr$^{-1}$ for $X$ and -1.1 mas yr$^{-1}$ for $Y$ direction. ; \citealt{Dehnen1998}). 
The systemic motion of G353 is expected to be negligible since rotating motion of G353 should be almost parallel to the sun, if we assume that G353 is moving along flat Galactic rotation.  
But if G353 has significant deviation from flat rotation, 
possible contribution can be up to 15 -- 30 km s$^{-1}$ (e.g., \citealt{Reid2009, Baba2009}). 
However, actual systemic motion seems to be enough small because no constant and systemic velocity vector has been seen for maser features in G353. 

\subsection{Maser acceleration}
Variations of $V_{\rmn{LSR}}$ during our monitoring is plotted in figure 5. 
Plotted $V_{\rmn{LSR}}$ values are flux density weighted means of relevant velocity channels in C50 and C70 (see table 3). 
Associated error bars show standard deviations from estimated means. These are good indicator of how maser emissions concentrate in the velocity domain. 
Filled squares and triangles correspond to single-dish and VLBI data, respectively. 
We also distinguish different peak flux densities by colour coding. 
Linear acceleration is evident in C70 during all detected epochs. 

\begin{figure*}
\includegraphics*[trim=0 0 0 0,scale=0.75]{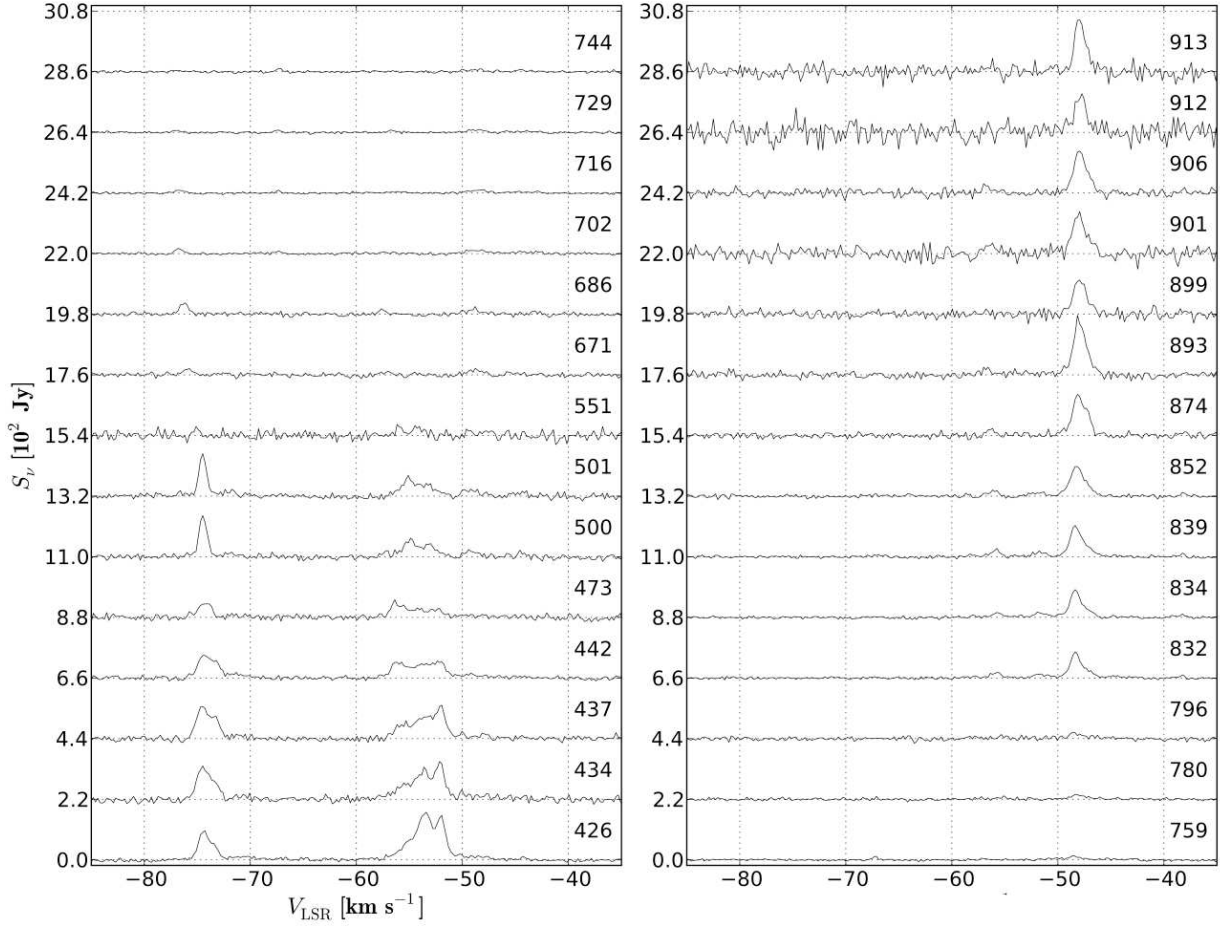}\\
\caption{Time variation of single-dish spectra. 
The vertical axis is in a hundred Jy unit. 
The numbers on far right show the relative day from Jan 1st, 2008. 
The systemic velocity of -5 km s$^{-1}$ and several weak high velocity components ($<$ -100 km s$^{-1}$) are out of the plot range. }
\end{figure*}

\begin{figure*}
\includegraphics*[trim=0 0 0 0,scale=0.45]{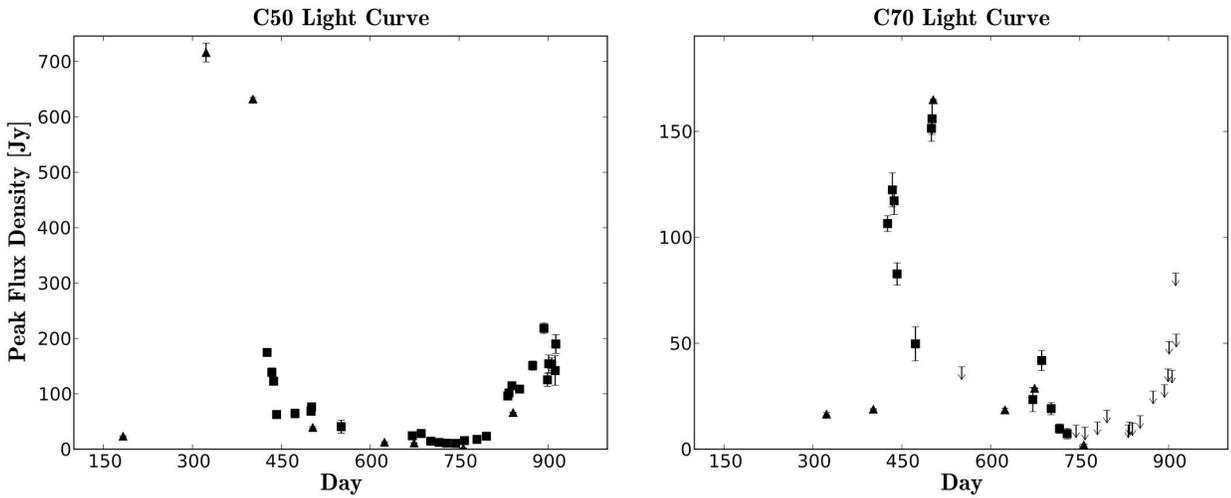} 
\caption{The light curve for C50 (left) and C70 (right). Filled squares and triangles correspond to single-dish and VLBI data, respectively. 
The horizontal and vertical axis shows the relative day from Jan 1st, 2008 and the peak flux density, respectively. 
The latter is sum of the flux densities of all detected spots in the peak velocity channel for VLBI data. 
Downward arrows present 3-$\sigma$ upper limit. }
\end{figure*}

\begin{figure*}
\begin{minipage}{140mm}
\includegraphics*[trim=0 0 0 0,scale=0.75]{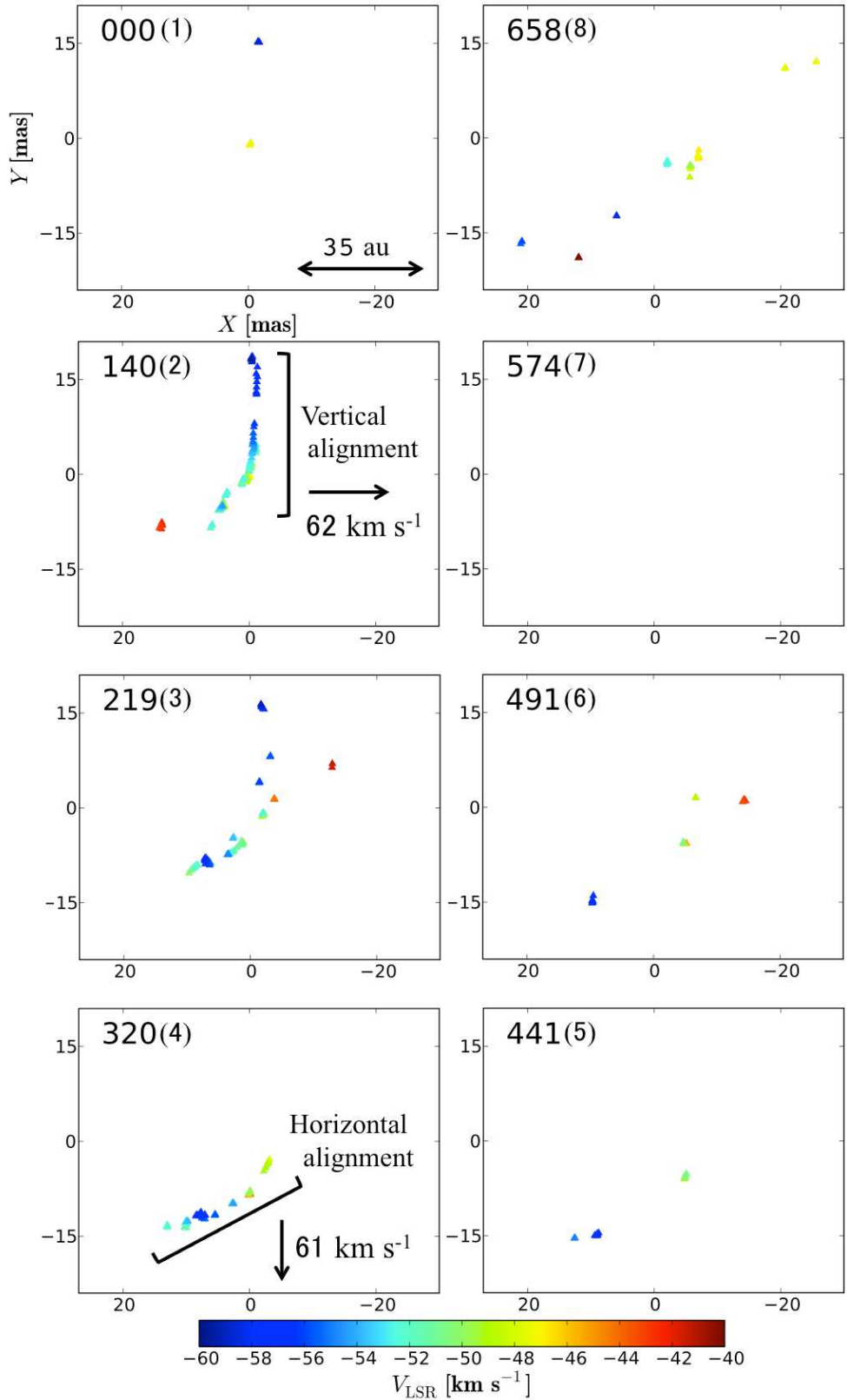}\\
\caption{Time variation of the prominent maser alignment around the brightest maser spot. 
Colour triangles indicate detected maser spots and their $V_{\rmn{LSR}}$. 
The numbers and associated parenthetic values in upper left side of each panel show the day offset from the first VLBI epoch and corresponding observing epochs, respectively. 
Two labelled velocities show estimated average proper motions assuming the distance of 1.7 kpc. }
\end{minipage}
\end{figure*}

\begin{figure}
\includegraphics*[trim=0 0 0 0,scale=0.44]{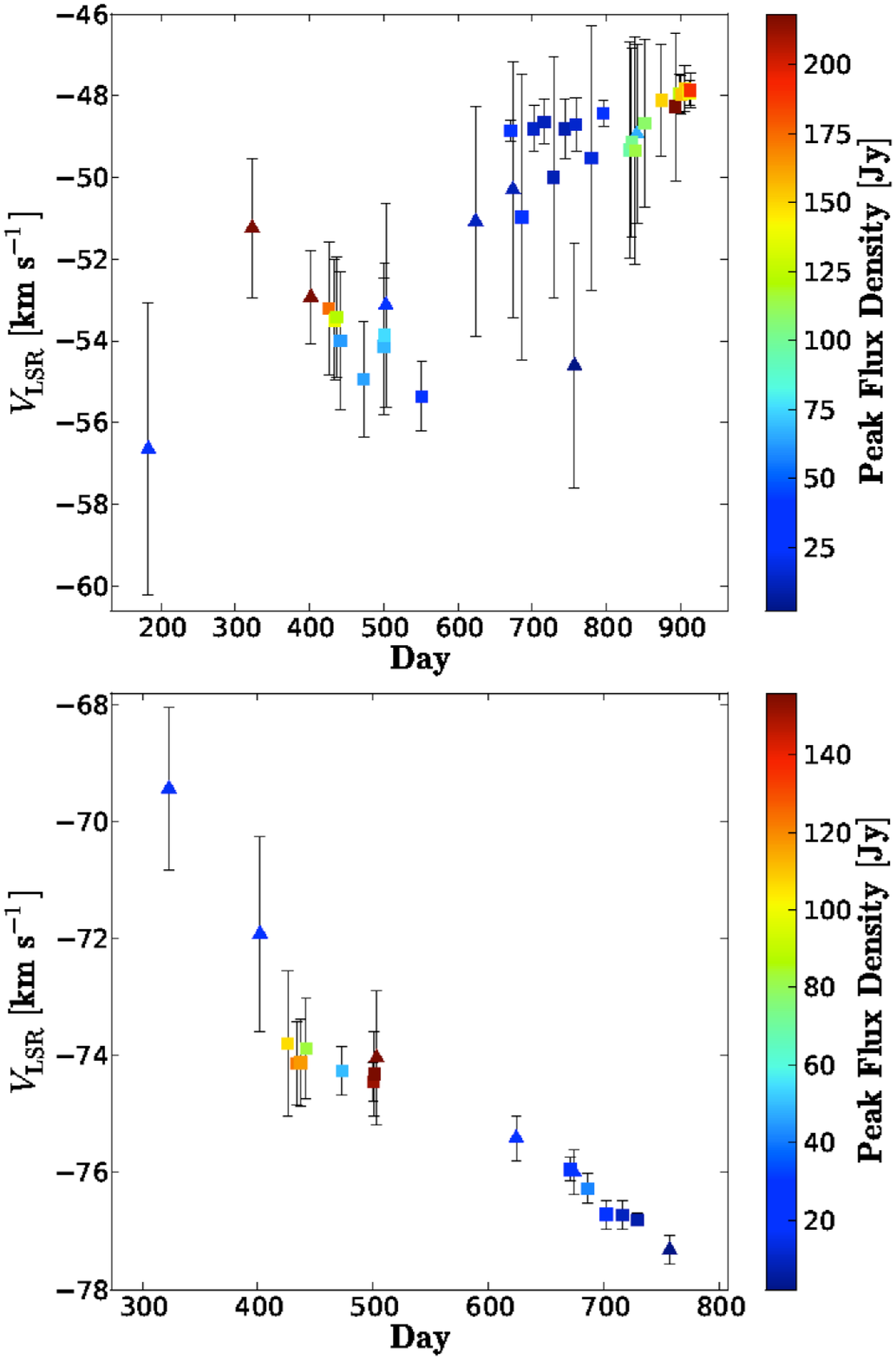}\\
\caption{Variations of the flux density weighted mean of $V_{\rmn{LSR}}$ for C50 (top) and C70 (bottom). 
The definitions of horizontal axis and markers are same as figure 4. The colour scale shows the peak flux densities. 
Error bars indicate standard deviations from the weighted means, which are measure of flux concentrations in the velocity domain. }
\end{figure}

Systematic acceleration is also seen for C50 only in the flare-up phase (day 300 -- 600), 
although these data points have relatively larger deviations compared to C70. 
Linearly fitted acceleration is -5.4 $\pm$ 1.2 km s$^{-1}$ yr$^{-1}$. 
This is consistent with the value of -4.8 $\pm$ 0.5 km s$^{-1}$ yr$^{-1}$ for C70 which is isolated from other components in both of velocity and spatial domains. 
The fact that two independent components have shown same acceleration suggests that 
it is not apparent acceleration discussed in \citet{Brand2003}, but real one caused by a momentum supply from a protostellar outflow or wind as is the case of W75N \citep{Hunter1994}. 

In addition to the acceleration, two discontinuous jumps of $V_{\rmn{LSR}}$ occurred in C50. 
The first jump was seen at the initial flare event ($\sim$ day 200) and the other one was synchronised with the run-out of the horizontal alignment ($\sim$ day 600). 
The initial jump clearly trace the flare-up of the maser alignment and the second jump is caused by attenuation and extinction of higher velocity ($\sim$ -56 km s$^{-1}$) components in C50.  

\section{Discussions}
\subsection{Outflow activity} 
\citet{Brand2003} has attributed one possible cause of long-term, grobal variation of H$_{2}$O masers to a variation of outflow activity itself. 
This seems to be applicable for G353, because it is difficult to fully attribute the systematic change of the shock front to internal fluctuations of coherent path. 
Synchronised acceleration of C50 and C70, which are separated $\sim$ 100 au in space and 20 km s$^{-1}$ in velocity, 
implies that subsequent momentum supply accelerates entire part of masing gas sheet, keeping or gradually losing coherence. 
If this is the case, intermittent maser flare of G353 is simply interpretable as episodic launching of an outflow. 
Expected launching time-scale is only $\sim$ 2 yr based on the interval between two flares in this case. 
This is well comparable with the typical time-scale of several radio jets \citep{Anglada1996}. 

\setcounter{table}{2}
\begin{table*}
\centering 
\begin{minipage}{120mm}
\caption{Parameters of the maser components}
\begin{tabular}{cccccc}\hline
Day & $\overline{V_{\rmn{LSR}}}^{a}$ & $\Delta\:V^{b}$ & Total Width & Peak Flux Density& Integrated Flux \\ 
& (km s$^{-1}$) & (km s$^{-1}$) & (km s$^{-1}$) & (Jy) & (Jy km s$^{-1}$) \\ \hline
\multicolumn{6}{c}{C50 single-dish data}\\ \hline
426 & -53.20 & 1.63 & 8.40 & 174.67 (3.74) & 617.91 (1.74) \\ 
434 & -53.49 & 1.48 & 5.67 & 138.75 (8.01) & 489.77 (3.67) \\ 
437 & -53.43 & 1.48 & 5.67 & 122.79 (6.54) & 412.05 (3.00) \\ 
442 & -54.00 & 1.70 & 5.88 & 62.63 (5.27) & 319.77 (2.40) \\ 
473 & -54.94 & 1.40 & 3.99 & 64.57 (8.01) & 204.19 (3.67) \\ 
500 & -54.13 & 1.67 & 4.62 & 68.64 (6.14) & 215.81 (2.80) \\ 
501 & -53.86 & 1.77 & 3.78 & 76.39 (7.41) & 212.60 (3.41) \\ 
551 & -55.36 & 0.85 & 0.42 & 40.73 (11.95) & 93.08 (5.48) \\ 
671 & -48.86 & 0.27 & 0.63 & 24.30 (5.74) & 33.59 (2.60) \\ 
686 & -50.97 & 3.49 & 1.05 & 28.38 (4.74) & 43.13 (2.14) \\ 
702 & -48.81 & 0.56 & 1.89 & 14.56 (2.87) & 37.99 (1.34) \\ 
716 & -48.64 & 0.55 & 1.68 & 12.29 (2.27) & 32.92 (1.07) \\ 
729 & -50.00 & 2.96 & 1.26 & 10.88 (2.47) & 24.57 (1.14) \\ 
744 & -48.81 & 0.74 & 0.84 & 10.55 (2.80) & 20.57 (1.27) \\ 
759 & -48.71 & 0.66 & 1.47 & 15.62 (2.47) & 36.79 (1.14) \\ 
780 & -49.53 & 3.24 & 1.89 & 17.76 (3.27) & 48.41 (1.47) \\ 
796 & -48.44 & 0.31 & 0.84 & 23.44 (5.14) & 66.10 (2.40) \\ 
832 & -49.34 & 2.65 & 5.67 & 96.08 (2.80) & 204.39 (1.34) \\ 
834 & -49.15 & 2.30 & 6.09 & 101.63 (3.27) & 211.07 (1.47) \\ 
839 & -49.35 & 2.79 & 6.72 & 114.65 (3.14) & 250.66 (1.47) \\ 
852 & -48.68 & 2.06 & 4.41 & 108.64 (4.27) & 235.30 (1.94) \\ 
874 & -48.11 & 1.37 & 2.94 & 151.04 (8.15) & 283.25 (3.74) \\ 
893 & -48.27 & 1.82 & 3.15 & 218.41 (9.15) & 368.71 (4.21) \\ 
899 & -47.95 & 0.46 & 1.89 & 125.33 (11.62) & 210.26 (5.34) \\ 
901 & -47.98 & 0.48 & 1.89 & 154.31 (15.96) & 275.03 (7.34) \\ 
906 & -47.82 & 0.56 & 2.31 & 153.64 (11.42) & 272.76 (5.27) \\ 
912 & -47.93 & 0.31 & 1.05 & 142.02 (26.71) & 234.97 (12.29) \\ 
913 & -47.88 & 0.43 & 1.68 & 189.97 (17.09) & 310.15 (7.81) \\ \hline
\multicolumn{6}{c}{C50 VLBI data}\\ \hline
183 & -56.65 & 3.57 & 2.52 & 23.50 (0.99) & 24.00 (0.56) \\ 
323 & -51.24 & 1.70 & 12.81 & 716.30 (17.01) & 2224.24 (14.28) \\ 
402 & -52.93 & 1.15 & 14.07 & 632.37 (2.86) & 969.45 (3.29) \\ 
503 & -53.13 & 2.50 & 9.24 & 39.42 (1.39) & 122.53 (1.51) \\ 
624 & -51.08 & 2.83 & 4.20 & 12.80 (0.94) & 21.95 (0.12) \\ 
674 & -50.30 & 3.13 & 4.62 & 11.26 (0.40) & 23.36 (0.05) \\ 
757 & -54.61 & 2.99 & 1.26 & 2.17 (0.37) & 2.10 (0.07) \\ 
841 & -48.94 & 2.19 & 6.51 & 66.42 (0.63) & 87.55 (0.12) \\ \hline
\multicolumn{6}{c}{C70 single-dish data}\\ \hline
426 & -73.79 & 1.24 & 4.20 & 106.50 (3.74) & 203.39 (1.94) \\ 
434 & -74.13 & 0.71 & 2.94 & 122.39 (8.01) & 233.3 (4.14) \\ 
437 & -74.12 & 0.75 & 3.15 & 117.25 (6.54) & 249.53 (3.34) \\ 
442 & -73.88 & 0.87 & 3.36 & 82.73 (5.27) & 202.32 (2.67) \\ 
473 & -74.26 & 0.41 & 1.47 & 49.74 (8.01) & 95.68 (4.14) \\ 
500 & -74.45 & 0.34 & 1.47 & 151.44 (6.14) & 168.47 (3.14) \\ 
501 & -74.31 & 0.72 & 1.89 & 155.98 (7.41) & 195.24 (3.81) \\ 
551 & -- & -- & -- & $<$ 35.86 (11.95) & -- \\ 
671 & -75.95 & 0.20 & 0.63 & 23.44 (5.74) & 35.12 (2.94) \\ 
686 & -76.28 & 0.25 & 0.84 & 41.87 (4.74) & 55.75 (2.40) \\ 
702 & -76.72 & 0.24 & 0.84 & 19.16 (2.87) & 11.02 (1.47) \\ 
716 & -76.72 & 0.25 & 0.84 & 9.68 (2.27) & 3.54 (1.14) \\ 
729 & -76.81 & 0.11 & 0.21 & 7.34 (2.47) & 7.41 (1.27) \\ 
744 & -- & -- & -- & $<$ 8.41 (2.80) & -- \\ 
759 & -- & -- & -- & $<$ 7.41 (2.47) & -- \\ 
780 & -- & -- & -- & $<$ 9.82 (3.27) & -- \\ 
796 & -- & -- & -- & $<$ 15.42 (5.14) & -- \\ 
832 & -- & -- & -- & $<$ 8.41 (2.80) & -- \\ 
834 & -- & -- & -- & $<$ 9.82 (3.27) & -- \\ 
839 & -- & -- & -- & $<$ 9.41 (3.14) & -- \\ 
852 & -- & -- & -- & $<$ 12.82 (4.27) & -- \\ \hline
\end{tabular}
\end{minipage}
\end{table*}%

\begin{table*}
\centering 
\begin{minipage}{120mm}
\contcaption{}
\begin{tabular}{cccccc}\hline
Day & $\overline{V_{\rmn{LSR}}}^{a}$ & $\Delta\:V^{b}$ & Total Width & Peak Flux Density& Integrated Flux \\ 
& (km s$^{-1}$) & (km s$^{-1}$) & (km s$^{-1}$) & (Jy) & (Jy km s$^{-1}$) \\ \hline
\multicolumn{6}{c}{C70 single-dish data}\\ \hline
874 & -- & -- & -- & $<$ 24.44 (8.15) & -- \\ 
893 & -- & -- & -- & $<$ 27.44 (9.15) & -- \\ 
899 & -- & -- & -- & $<$ 34.85 (11.62) & -- \\ 
901 & -- & -- & -- & $<$ 47.88 (15.96) & -- \\ 
906 & -- & -- & -- & $<$ 34.25 (11.42) & -- \\ 
912 & -- & -- & -- & $<$ 80.13 (26.71) & -- \\ 
913 & -- & -- & -- & $<$ 51.28 (17.09) & -- \\ \hline
\multicolumn{6}{c}{C70 VLBI data}\\ \hline
323 & -69.44 & 1.40 & 4.41 & 16.60 (0.68) & 38.03 (0.51) \\ 
402 & -71.92 & 1.67 & 7.35 & 18.90 (0.22) & 65.93 (0.28) \\ 
503 & -74.04 & 1.14 & 6.51 & 165.00 (0.26) & 175.37 (0.44) \\ 
624 & -75.41 & 0.38 & 1.89 & 18.60 (0.80) & 19.77 (0.23) \\ 
674 & -75.99 & 0.38 & 2.10 & 28.80 (0.15) & 24.79 (0.23) \\ 
757 & -77.32 & 0.25 & 0.63 & 2.04 (0.34) & 1.08 (0.11) \\ \hline
\multicolumn{6}{l}{$^a$ Flux density weighted mean.}\\ 
\multicolumn{6}{l}{$^b$ Standard deviations from the weighted means.}\\ 
 \multicolumn{6}{l}{All parenthetic values indicate associated errors.}\\
\end{tabular}
\end{minipage}
\end{table*}%

The characteristic velocity of $\sim$ 500 km s$^{-1}$, which is estimated from the spatial scale of maser distribution ($\sim$ 200 au) divided by the time-scale ($\sim$ 2yr), 
is also consistent with the proper motion velocity of radio jets associated with several high mass YSOs (e.g., \citealt*{Marti1998, Curiel2006}). 
Thus, observed variation can be caused by standard radio jet in high mass star-forming region. 
The relation between the expected velocity of a jet and the 3D velocity of masers ($\sim$ 100 km s$^{-1}$) is roughly same as the case of CepA HW2 region \citep{Torrelles2011}, 
where velocities of radio jet and relatively wide angle maser flow is 500 and 70 km s$^{-1}$, respectively. 
G353 is thought to be a possible candidate of high mass YSO with a radio jet activity through these reasons. 

The momentum rate of maser flow is estimated as 1.1 $\times$ 10$^{-3}$ $M_{\sun}$ km s$^{-1}$ yr$^{-1}$ based on the acceleration of $\sim$ 5 km s$^{-1}$ yr$^{-1}$. 
Here, we assumed that the masing gas sheet has uniform thickness of 1 au, i.e., typical size of maser feature. 
Total area of the sheet is evaluated as 200 $\times$ 200 au$^{2}$ from the extent of maser distribution. 
Adopted number density of molecular hydrogen is 10$^{9}$ cm$^{-3}$ that is also typical value for a H$_{2}$O maser feature. 
We note that assumed volume is, and hence, the momentum rate is only a lower limit, because a mass fraction of masing gas sheet to total shocked gas should be small \citep{Elitzur1989}. 
Estimated lower limit is clearly larger than a typical values of low mass star-formation ($\sim$ 10$^{-5}$ $M_{\sun}$ km s$^{-1}$ yr$^{-1}$; \citealt{Arce2006}) and consistent with that of high mass star-formation (e.g., \citealt{Arce2007} for comparison). 
If we adopt a jet velocity of 500 km s$^{-1}$, it is immediately converted to the outflow rate of 2.2 $\times$ 10$^{-6}$ M$_{\sun}$ yr$^{-1}$. 
We finally get the lower-limit accretion rate of 10$^{-5}$ -- 10$^{-4}$ M$_{\sun}$ yr$^{-1}$, 
where we have adopted typical fraction between outflow rate over accretion rate (1 -- 10 \%) in a case of MHD driven flow (e.g., \citealt{Pelletier1992, Machida2006}). 

Although we have considered pure MHD jet (e.g., \citealt*{Machida2008}) here, strong radiation from the host YSO should also play an important role for outflow launching in high mass star-fomation. 
If the host YSO is enough young and an accretion rate is enough high, accretion luminosity is thought to be significant or even dominant in total flux \citep{Hosokawa2010}. 
Outflow variations may directly reflect variations of accretion rate onto the protostellar surface in this case. 
If this is true, variation pattern of the H$_{2}$O maser can be important clue for accretion mechanism at the innermost part. 
Further discussion about this requires exact stellar parameters in IR wavelength. 

Another explanation for intermittent maser flare is a precessing jet, where we can observe maser emission only in a certain fraction of precession period because of beaming effect. 
If simple binary interaction which has rotating period of 2 yr is considered, 
expected binary separation is over 3 au with a total mass of 10 M$_{\sun}$. 
By contrast, if we assume escape velocity at he launching point as typical velocity of a jet, expected launching radius is estimated as follows, 
\begin{eqnarray*}
R_{\rmn{launch}} = 0.04 \times \left(\frac{M_{\rmn{object}}}{10\:M_{\rmn{\sun}}} \right) \left(\frac{V_{\rmn{jet}}}{500\:\rmn{km}\:\rmn{s}^{-1}}\right)^{-2} (\rmn{au}), \nonumber
\end{eqnarray*}
here $M_{\rmn{object}}$ and $V_{\rmn{jet}}$ indicate mass of the driving source and velocity of launched jet, respectively. 
It is well smaller than the expected binary separation and this scenario seems to be possible. 
High-resolution image of molecular lobe obtained with upcoming Atacama Large Millimeter / submillimeter Array (ALMA) will be helpful to distinguish two scenarios. 
Direct detection of precessing trajectory of a jet is also the best way to test them. 
This can be achieved by high-sensitivity observation of radio continuum or IR line emissions, as the case of IRAS16547-4247 (\citealt{Brooks2003, Rodriguez2005}). 

\subsection{Driving source}
The H$_{2}$O maser distribution is completely inside the error circle of the position of class II CH$_{3}$OH maser reported by CP08. 
Since there is no detectable OH maser emission (e.g., CP08; \citealt{Breen2010}), 
these masers are thought to be excited by identical high mass protostellar object. 
If the H$_{2}$O masers in G353 is actually excited by a radio jet, 
such a close location ($<$ a few hundred au) of two types of masers and the radio jet is also similar to the cases of several high mass protostellar objects. 
Besides CepA HW2 \citep{Torstensson2011}, $IRAS$ 20126+4104 is the best example of this kind of source (\citealt{Hofner2007, Moscadelli2011}). 

H$_{2}$O masers in $IRAS$ 20126+4104 are obviously excited in front of the radio knot which emerges from the nearly edge-on disc (e.g., \citealt{Edris2005, Cesaroni2005}). 
The intriguing feature of $IRAS$ 20126+4104 is asymmetry of the radio knot and H$_{2}$O maser in \citet{Hofner2007}. 
In spite of overall bipolar structure of outflow verified by radio and IR line observations \citep{Su2007, Garatti2008}, 
the radio knot and the most of H$_{2}$O masers have been detected only in blue-shifted side. 
The non-negligible blue-shift dominance of H$_{2}$O maser is also evident in long-term single-dish study \citep{Felli2007}. 
Their light curve, furthermore, has shown weak intermittent variation of 2 -- 3 yr scale alike G353. 

These facts strongly support the interpretation that G353 is a candidate of high mass protostellar object with intermittent maser activity caused by a protostellar jet, as proposed in CP08.  
If this is the case, G353 can be a good target of direct imaging of the accretion disc, because of the possible pole-on geometry of the disc-jet system (e.g., \citealt{Kraus2010}). 

\section{Summary and Conclusion}
Our VLBI and single-dish monitoring of 22 GHz H$_{2}$O maser emission from the high mass young stellar object G353.273+0.641 
with VERA and Tomakamai 11-m radio telescope has detected two times of maser flares. 
The eight epochs of VLBI data have revealed that these flares have been accompanied by structural change of the prominent shock front traced by H$_{2}$O maser alignments. 
We have detected only blue-shifted maser emissions and all maser features have been distributed within very small area (200 $\times$ 200 au$^{2}$) in spite of wide velocity range ($>$ 100 km s$^{-1}$). 
These facts suggest relatively pole-on geometry of the disc-jet system. 

The light curve shows notably intermittent variation. This suggests a possibility that the H$_{2}$O masers in G353.273+0.641 are excited by episodic radio jet. 
The interval of two flares is about two years and well comparable to typical time-scale of a radio jet. 
The characteristic velocity of 500 km s$^{-1}$, which is estimated from the time and spatial scale, 
is also consistent with the velocity of the radio jets associated with several high mass young stellar objects. 
These facts also suggest that the H$_{2}$O maser in G353.273+0.641 is excited in a strong shock of a radio jet. 
The precessing jet may be also able to explain intermittent maser flare and this will be tested by 
high-resolution imaging of molecular lobe or direct detection of precessing trajectory of a jet. 

Two strong velocity components of C50 (-53 $\pm$ 7 km s$^{-1}$) and C70 (-73 $\pm$ 7 km s$^{-1}$) have shown synchronised linear acceleration of the flux weighted $V_{\rmn{LSR}}$ values ($\sim$ -5 km s$^{-1}$ yr$^{-1}$) during the flare phase. 
The acceleration can be converted to the lower-limit momentum rate of 1.1 $\times$ 10$^{-3}$ M$_{\sun}$ km s$^{-1}$ yr$^{-1}$. 
If we use a jet velocity of 500 km s$^{-1}$, expected lower limit of accretion rate is 10$^{-5}$ -- 10$^{-4}$ M$_{\sun}$ yr$^{-1}$ in the case of MHD driven jet. 

The co-existence of class II CH$_{3}$OH maser at 6.7 GHz and absence of any detectable OH masers, as pointed-out in CP08, indicate that the driving source is a high mass protostellar object. 
If H$_{2}$O masers are actually excited by a radio jet, 
the close location ($<$ a few hundred au) of two types of masers and the radio jet is also similar to the cases of several high mass protostellar objects. 
$IRAS$ 20126+4104 especially has H$_{2}$O masers excited by the radio jet and 
quite resembles G353.273+0.641 in blue-shift dominant spectra and short-term (2 -- 3 yr) intermittent variation of the maser. 
We conclude that G353 is a candidate of high mass protostellar object alike $IRAS$ 20126+4104, as proposed in CP08. 
The possible pole-on geometry of disc-jet system can be suitable for direct imaging of the accretion disc in this case. 

\section*{Acknowledgments}
We would like to thank all the member of VERA and Hokkaido University for their assistance in observations and data analyses. 
We also thank the referee for helpful comments. 
This work was financially supported by the Grant-in-Aid for the Japan Society for the Promotion of Science Fellows (K. M.).

\label{lastpage}

\end{document}